\documentclass[aps,nofootinbib,superscriptaddress,twocolumn,prd,10pt]{revtex4-1}

\usepackage{graphicx}
\usepackage{amsmath,amssymb}
\usepackage{hyperref}
\usepackage{braket}
\usepackage{booktabs}
\usepackage{subfigure}
\usepackage{float}
\usepackage[dvipsnames]{xcolor}
\usepackage{mathrsfs}
\usepackage{comment}
\usepackage{bm}
\usepackage[normalem]{ulem}
\usepackage{sistyle}
\SIthousandsep{,}

\hypersetup{
    colorlinks=true,
    linkcolor=red,
    citecolor=blue,
}

\newcommand{\rmx}{\mathrm{X}}

\usepackage{natbib}
\usepackage[utf8]{inputenc}
\usepackage{setspace}

\usepackage[margin=1in]{geometry}
\usepackage{graphics}      

\newcommand{\fnl}{\ensuremath{f_\mathrm{NL}}}

\newcommand{\planck}{\textit{Planck}}
\newcommand{\fsky}{\ensuremath{f_\mathrm{sky}}}

\newcommand{\lmin}{\ensuremath{\ell_\mathrm{min}}}

\begin{document}

\title{The Square Kilometer Array as a Cosmic Microwave Background Experiment}

\author{David Zegeye}
\email{dzegeye@uchicago.edu}
\affiliation{Department of Astronomy \& Astrophysics, The University of Chicago, Chicago, IL 60637, USA}
\affiliation{Kavli Institute for Cosmological Physics, The University of Chicago, Chicago, IL 60637, USA}

\author{Thomas Crawford}
\affiliation{Department of Astronomy \& Astrophysics, The University of Chicago, Chicago, IL 60637, USA}
\affiliation{Kavli Institute for Cosmological Physics, The University of Chicago, Chicago, IL 60637, USA}

\author{Jens Chluba}
\affiliation{Jodrell Bank Centre for Astrophysics, Alan Turing Building, University of Manchester, Manchester M13 9PL}

\author{Mathieu Remazeilles}
\affiliation{Instituto de Física de Cantabria (CSIC-UC), Avda. de los Castros s/n, 39005 Santander, Spain}

\author{Keith Grainge}
\affiliation{Jodrell Bank Centre for Astrophysics, Alan Turing Building, University of Manchester, Manchester M13 9PL}

\begin{abstract}

Contemporary cosmic microwave background (CMB) experiments typically have observing bands covering the range 20 - 800\;GHz. Certain science goals, including the detection of $\mu$-type distortions to the CMB spectrum and the characterization of low-frequency foregrounds, benefit from extended low-frequency coverage, but the standard CMB detector technology is not trivially adaptable to radio wavelengths. 
We propose using the upcoming Square Kilometer Array (SKA) as a CMB experiment, exploiting the immense raw sensitivity of SKA, in particular in single-dish mode, to measure medium-to-large-angular-scale modes of the CMB at radio wavelengths. As a worked example, we forecast the power of SKA combined with the upcoming LiteBIRD CMB space mission to constrain primordial non-Gaussianity through measurements of the correlation between anisotropies in the CMB $\mu$-distortion, temperature, and $E$-mode polarization fields. We find that adding SKA data significantly improves the constraints on \fnl, even {for} spatially varying low-frequency {foregrounds}.
\end{abstract}
\maketitle

\section{Introduction}

The Square Kilometre Array (SKA, \cite{Huynh:2013aea}) is a planned array of radio telescopes, aiming to probe the Universe with unprecedented resolution and sensitivity, primarily through observations of the neutral hydrogen 21-cm hyperfine transition \cite{Braun:2015zta}.
SKA 21-cm data have the potential to make groundbreaking contributions to our understanding of cosmology and astrophysics.
More fundamentally, SKA will make maps of the sky at many frequencies (from 50\;MHz to 15\;GHz) with high spectral resolution. Contemporary
experiments designed to measure the cosmic microwave background (CMB) typically observe only down to $\sim$30\;GHz, limited primarily by the challenge of adapting standard CMB detector technology to long wavelengths \cite{Benford:2009zz}. While this is sufficient for many science goals, certain applications of CMB data suffer from the lack of lower-frequency information, in particular the search for $\mu$-type spectral distortions \cite{Abitbol:2017vwa,CMB-S4:2023zem} and the characterization of the Galactic synchrotron signal \cite{Krachmalnicoff:2015xkg}. 

We propose using SKA data not to trace neutral hydrogen but as a low-frequency CMB experiment. SKA is well-matched and complementary to current and planned CMB experiments in several ways:
\begin{itemize}
    \item \textit{Raw sensitivity}: The expected noise in SKA wide survey maps, in CMB units, is below 
    10 $\mu$K-arcmin for SKA-1 (and even lower for SKA-2) for sufficiently wide frequency bands, which approaches the projected map depths for, e.g., CMB-S4 \cite{Abazajian:2019eic} and LiteBIRD \cite{LiteBIRD:2022cnt}.
    \item \textit{Range of angular scales}: Cosmologically relevant information in the CMB is concentrated at angular scales of roughly tens of degrees down to several arcminutes (multipoles $10 \lesssim \ell \lesssim 3000$). The SKA will operate in two modes: Interferometric mode, which will access the high-$\ell$ part of this range, and single-dish mode, which will recover the low-$\ell$ part of this range. 
    \item \textit{Spectral resolution}: The ability to channelize the SKA data will provide flexibility in treating foregrounds. If foregrounds are more complex than anticipated, SKA data can be analyzed at high spectral resolution, with some noise penalty; if foregrounds turn out to be relatively simple, SKA data can be combined into wider bands to gain sensitivity.
\end{itemize}
While there have been previous suggestions of using SKA as a CMB experiment \cite{Subrahmanyan:2002aj}, to our knowledge, no detailed forecasts exist.

In the worked example of constraining primordial non-Gaussianity at effective scales of $k\approx740 \; \textrm{Mpc}^{-1}$ from correlations between $\mu$ distortion and CMB temperature and $E$-mode polarization \cite{Pajer:2012vz,Ganc:2012ae}, we find that the combination of SKA with planned CMB missions can significantly improve constraints, even in the case of spatially varying foreground properties. 
This example can likely be generalized to other CMB science cases, such as detecting $B$-modes generated from primordial gravitational waves during inflation.

\section{Theory Background}
\label{sec:theory}

$\mu$-distortions are generated when energy injection into the early-universe photon-baryon plasma cannot be efficiently thermalized although Comptonization is still efficient. This leads to a Bose-Einstein distribution $n(\nu)=\left[e^{h \nu /\left(k_{B} T\right)+\mu(\nu)}-1\right]^{-1}$ \cite{Sunyaev:1970eu,Burigana:1991eub, Hu:1992dc, Chluba:2011hw}. 
The frequency-dependent chemical potential, $\mu(\nu)$, is approximately constant above 500\;MHz, but exponentially decays at lower frequencies due to double Compton scattering and Bremsstrahlung 
\cite{Zeldovich:1969ff, Chluba:2013kua}. A $\mu$-type distortion is formed until redshift $z \simeq 5\times10^4$, after which a $y$-type distortion is produced \cite{Zeldovich:1969ff,Chluba:2016bvg}. 

In standard cosmology, the primary energy injection that generates $\mu$-distortions is diffusion damping of the primordial power spectrum $P_{\zeta}(k)$ at small scales \cite{Sunyaev:1970eu, Daly:1991uob, Hu:1994bz,Chluba:2012gq}. The average $\mu$-distortion from diffusion damping is related to the power spectrum as: 
\begin{eqnarray}
\langle\mu \rangle \propto \int_{0}^{\infty} d k\int_{0}^{\infty}{d}z \frac{ P_\zeta(k) k^4}{2 \pi^2}     \frac{d k_\mathrm{D}^{-2}}{dz}  \mathrm{e}^{\frac{-2k^{2}}{k_\mathrm{D}^{2}}}\mathcal{J}_{\mu}(z) , 
\end{eqnarray}
where $k_\mathrm{D}$ is the damping scale, and $\mathcal{J}_\mu$ is the time window function for $\mu$ distortions \cite{Chluba:2012we,Chluba:2016aln}.

If inflation
is driven by a single field initially in a Bunch–Davies vacuum, with no additional interactions, 
the resulting primordial curvature perturbations $\zeta(\vec{k})$ are purely Gaussian, and the power spectrum is statistically isotropic. The presence of additional fields and interactions during inflation introduces higher-order correlations between curvature perturbations, introducing statistical anisotropy, or non-Gaussianity. In the scenario where a long-wavelength curvature mode $k_L$ is correlated with much smaller modes $k_S$, the long mode modulates the power spectrum at small scales. In Fourier space, this correlation, represented by a triangle configuration of momentum with two sides given by $k_S$ and one by $k_L$, is known as the \textit{squeezed-limit bispectrum}. In real space, this results in a spatially varying small-scale power spectrum, thus inducing \textit{anisotropies} in the $\mu$-distortion, which can be correlated with CMB modes \cite{Pajer:2012vz, Ganc:2012ae, Emami2015muT, Khatri2015muT, Chluba:2016aln, Dimastrogiovanni:2016aul, Orlando:2021nkv, McCulloch:2024hiz}. 

Following \cite{Cabass:2018jgj, CMB-S4:2023zem}, the angular cross-power spectrum 
of $\mu$ anisotropies and CMB modes is
\begin{equation}
C_{\ell}^{\mu X}  = \frac{24 \langle\mu\rangle}{5 \pi } \fnl  \int_{0}^{\infty} \mathrm{d} k  P_{\zeta}(k) k^2 \Delta_{\ell}^{\mu}(k) \Delta_{\ell}^{X}(k) ,
\label{eqn:mux}
\end{equation}
where $\fnl$ parameterizes the amplitude of the correlation, $X\in\{T,E\}$  and $\Delta_\ell^X$ is the corresponding transfer function, calculated {using} CAMB \cite{Lewis:1999bs}.
For the $\mu$-anisotropy transfer function, we {use $\Delta_\ell^\mu\approx {\rm e}^{-k^{2}/\left(q_{\mu,\mathrm{D}}^{2}\left(z_{\mathrm{rec}}\right)\right)} j_{\ell}(k\eta_{0}-k\eta_{\mathrm{rec}})$ \cite{Pajer:2012qep, Chluba:2016aln, Ravenni:2017lgw, Cabass:2018jgj}},
with damping scale $q_{\mu, \mathrm{D}}\left(z_{\mathrm{rec}}\right) \approx 0.11 \, \mathrm{Mpc}^{-1}$ \cite{Chluba:2016aln}.
\section{Methods}
\label{sec:methods}
\subsection{Fisher Matrix}
Following \cite{Remazeilles:2018kqd,Remazeilles:2021adt,CMB-S4:2023zem}, we forecast constraints on $\fnl$ from measurements of the $\mu \times T$ and $\mu \times E$ cross-spectra $C_{\ell}^{\mu T}$ and $C_{\ell}^{\mu E}$ using a Fisher-matrix approach. For our single-parameter model (given by Eq.~\ref{eqn:mux} with fixed value for $\langle \mu \rangle$ and free parameter \fnl), the Fisher ``matrix" is a scalar. Here, the shape of the spectrum is fixed, and \fnl\ controls the overall amplitude, yielding an analytical expression for the projected $1 \sigma$ uncertainty on \fnl:
\begin{eqnarray}
&& \sigma(\fnl) = \bigg( \sum_{\ell=\lmin}^{\ell_{\max }}\frac{(2 \ell+1) \fsky}{C_{\ell}^{\mu \mu}\left[C_{\ell}^{T T} C_{\ell}^{E E}-\left(C_{\ell}^{T E}\right)^{2}\right]} \times \\
\nonumber && \bigg [ C_{\ell}^{T T}\left( C_{\ell}^{\mu E}|_{\fnl=1} \right)^{2}+C_{\ell}^{E E}\left(C_{\ell}^{\mu T}|_{\fnl=1}\right)^{2}- \\
\nonumber && 2 C_{\ell}^{T E}  C_{\ell}^{\mu T}|_{\fnl=1} C_{\ell}^{\mu E}|_{\fnl=1} \bigg ] \bigg)^{-1/2}.
\label{eqn:fulllike}
\end{eqnarray}
As in \cite{CMB-S4:2023zem}, we assume measurements of the primary CMB power spectra $C_{\ell}^{T T}$, $C_{\ell}^{T E}$, and $C_{\ell}^{E E}$ 
are signal-dominated at the scales of interest to this work, while $C_{\ell}^{\mu \mu}$ will be dominated by noise and foregrounds. Similarly to \cite{CMB-S4:2023zem, Zegeye:2023pbl}, we model the contribution to the band-band $\ell$-space covariance matrix $\mathbf{C}^{\mathrm{ij}}_\ell$ from noise as 
\begin{equation}
    \mathbf{C}^{\mathrm{ij,N}}_\ell = (N^i)^2 
    e^{\ell^2 \theta_i^2  / (8 \ln(2))} \delta_{ij},
\end{equation}
where $N^i$ is the white noise level in the map from band $i$, $\theta_i$ is the beam FWHM in band $i$, and from foreground $\rmx$ as
\begin{equation}
    \mathbf{C}_\ell^{ij,\rmx} = \sqrt{C_\rmx(\ell,\nu_i) C_\rmx(\ell,\nu_j)}.
\label{eq:correlatednoise}
\end{equation}
The $\mu \times \mu$ covariance matrix is then given by 
\begin{eqnarray}
    {C_{\ell}^{\mu \mu}} =  \sum_{ij} \boldsymbol{w}_{0i}\mathbf{C}^{\mathrm{ij}}_{_\ell} \boldsymbol{w}_{0j},
\end{eqnarray}
where $\boldsymbol{w}$ are the weights used to produce a $T$-free $\mu$ map from the individual band maps \citep[e.g.,][]{Remazeilles:2018kqd, CMB-S4:2023zem}.

\subsection{Foregrounds}
\label{sec:fg}

We closely follow the Galactic and extragalactic foreground treatment of \cite{CMB-S4:2023zem, Zegeye:2023pbl}. 
For Galactic foregrounds, we consider synchrotron, dust, and anomalous microwave emission. For extragalactic foregrounds, we consider the thermal Sunyaev-Zeldovich (tSZ) effect, 
the cosmic infrared background (CIB), and synchrotron-emitting active galactic nuclei.
We model each foreground component as a power law in both frequency and $\ell$ space, adopting the values for power-law indices and ``Wide survey" amplitude values in \cite{Zegeye:2023pbl}. As indicated by Eq.~\ref{eq:correlatednoise}, we assume each foreground component is 100\% correlated across frequency bands. 

Two low-frequency foregrounds that we neglect here are Galactic free-free emission and a potential source related to the low-frequency excess emission reported by the ARCADE team \cite{Fixsen:2009xn}. For free-free emission, we note that the free-free templates in the \planck\ Sky Model \cite{Delabrouille:2012ye} and PySM \cite{Thorne:2016ifb} are dominated by point-like sources, either real HII regions (which can be masked) or contamination from extragalactic radio sources (which are already in our foreground model). If the ARCADE excess is a true astrophysical background that clusters at some level \citep[][for discussion]{Singal:2017jlh,Singal:2022jaf}, it will contaminate the searches discussed here, but also make a new scientific target that SKA will be well-positioned to constrain.

\subsection{Moment Expansion}

Our foreground treatment implicitly assumes 
constant amplitude and spectral indices across the sky. While this roughly holds for extragalactic foregrounds, galactic foregrounds have significant spatial variations across the sky. 
When attempting to characterize extremely faint signals such as primordial $B$ modes or $\mu$ distortions, the isotropic approximation can lead to significant biases on the parameters of interest \cite{Remazeilles:2015hpa,Remazeilles:2020rqw}.

At the Fisher level, this can be treated using a moment expansion formalism \cite{Chluba:2017rtj}, which we apply to account for spatially varying spectral indices for dust and synchrotron. We do not consider variations in their amplitude, given that the isotropic signal is so bright the covariance matrix effectively marginalizes out the corresponding spectral energy distribution (SED) \cite{Bond:1998zw}. 
We also ignore spatial variations in the dust temperature $T_\mathrm{d}$ since that is nearly degenerate with changes in amplitude in the frequency bands of interest to $\mu$-distortions.

In this work, we consider the two largest contributions to the covariance from auto- or cross-power spectra of moment terms: $1 \times 1$ and $0 \times 2$, in addition to $0 \times 0$, the contribution from the component with the mean spectral behavior. We follow \cite{Azzoni:2020hpw} 
and calculate these as:
\begin{equation}
\begin{aligned}
&C_{\ell,\nu_1,\nu_2}^{\rmx,1\times1} =\log{\left(\frac{\nu_1}{\nu_0}\right)}\log{\left(\frac{\nu_2}{\nu_0}\right)}\times\\
&\left.\sum_{\ell_{1}\ell_{2}}\frac{(2\ell_{1}+1)(2\ell_{2}+1)}{4\pi}\left(\begin{array}{ccc}\ell&\ell_{1}&\ell_{2}\\0&0&0\end{array}\right.\right)^{2}C_{\ell_1,\nu_1,\nu_2}^{\rmx,0\times0}C_{\ell_{2}}^{\beta_{\rmx}},  \\
&C_{\ell,\nu_1,\nu_2}^{\rmx,0\times2} =\frac{\sigma_{\beta_{\rmx}}^{2}}2\left[\log{\left(\frac{\nu_1}{\nu_0}\right)}^2+\log{\left(\frac{\nu_2}{\nu_0}\right)}^2\right]C_{\ell,\nu_1,\nu_2}^{\rmx,0\times0}, \\
&\sigma_{\beta_\rmx}^2\equiv\sum_{\ell}\frac{2\ell+1}{4\pi}C_{\ell}^{\beta_\rmx}, \quad C_\ell^{\beta_\rmx}=B_\rmx\left(\frac\ell{\ell_0}\right)^{\gamma_\rmx},
\end{aligned}
\end{equation}
where $\rmx\in\{\mathrm{dust},\mathrm{synchrotron}\}$, $\sigma_{\beta_{\rmx}}^{2}$ is the variance in spectral index across the sky, and $C_\ell^{\beta_\rmx}$ is the angular power spectrum of the spectral index variations, assumed to be a power law in $\ell$ with amplitude $B_\rmx$ and index $\gamma_\rmx$.

\vspace{-3mm}
\subsection{Survey specifications}
We forecast our constraints on \fnl\ from $\mu \times T$ and $\mu \times E$ based on Phase 1 of the planned SKA Observatory, augmented with data from existing or upcoming ``traditional'' CMB experiments.
Phase 1 of SKA is under construction and encompasses two arrays: a ``low-frequency'' array (SKA1-LOW) in Australia that will observe from 50-350\;MHz, and a ``mid-frequency'' array (SKA1-MID) in South Africa observing at 350\;MHz to 15.4\;GHz, with the goal of extending to 24\;GHz.\footnote{skao.int/en/science-users/118/ska-telescope-specifications}
Since the $\mu$-distortion SED peaks below 1\;GHz, both arrays have the potential to isolate and identify $\mu$-distortions from foregrounds. 
Given the challenges of modeling the foreground behavior at low frequencies, we only consider the SKA1-MID array.

One of two proposed cosmology surveys for SKA1-MID is a ``wide survey'' of the southern sky \cite{SKA:2018ckk}. In the default configuration for the wide survey, each telescope operates in single-dish mode, acting as an individual detector. 
The highest signal-to-noise ratio for $\mu \times T$ and $\mu \times E$ is at low $\ell$, so single-dish mode is the preferred configuration for our purposes. 

The angular resolution of each dish at center frequency $\nu_i$ is $\theta_i={1.22c} ({D_\text{dish}  \nu_i})^{-1}$,
where $D_\text{dish}\approx14$ meters is the dish diameter, and $c$ is the speed of light. The noise for SKA1-MID in single-dish mode at frequency channel $i$ is

\begin{equation}
 N^{i}=\sqrt{\frac{T_{\mathrm{sys,i}}^{2} S}{N_{\mathrm{d}}  N_{\mathrm{b}} N_{\mathrm{p}} t \Delta \nu } },
 \end{equation}
where $S=\num{20000} \textrm{ deg}^2$ is the area of the wide survey; $t=10^4$ hours is the total on-sky time; $\Delta \nu$ is the channel bandwidth; $N_\text{d}$ is the number of dishes, which for SKA1-MID is 197 \cite{SKA:2018ckk}; $N_\text{b}$ is the number of simultaneous observing beams, which is 1 for SKA1-MID; and $N_\text{p}$ is the number of independent Stokes I measurements, which is 2. 
$T_\textrm{sys}$ is the system temperature, with contributions from receiver noise $T_\textrm{rcvr}$, spill-over $T_\textrm{spl}\approx3$K,  the CMB (assuming negligible contribution from spectral distortions) $T_\text{CMB}=2.73$K, and Galactic emission $T_\textrm{gal}\approx25\textrm{K}(408 \; \textrm{MHz}/\nu_i)^{2.75}$ \cite{SKA:2018ckk}.
{Above $1~{\rm GHz}$},  we assume $T_\text{rcvr} = 7.5$K and {below} we set $T_\text{rcvr} = 15\textrm{K}+30\textrm{K}\left(\frac{\nu_i}{1 \; \textrm{GHz}}-0.75\right)^2$ \cite{SKA:2018ckk}. We neglect the atmospheric contribution, noting that at Karoo, South Africa it will be subdominant in the total sky contribution compared to $T_\text{CMB}$ and $T_\textrm{gal}$. \footnote{skao.int/sites/default/files/documents/Anticipated
\%20Performance\%20of\%20the\%20SKA.pdf}

The SKA1-MID receiver bands will be sub-divided into \num{65000} frequency channels, useful for identifying and removing RFI and other systematics, and can then be combined into wider bands.
For frequencies below 2.5\;GHz, we set $\Delta\nu = 100$\;MHz to better isolate the peak of the $\mu$ spectrum, while for frequencies above 2.5\;GHz we set $\Delta \nu = 1$\;GHz to obtain lower noise for improved calibration off of the CMB. 

\begin{figure}[t!]
\centering
\includegraphics[width=1\columnwidth]{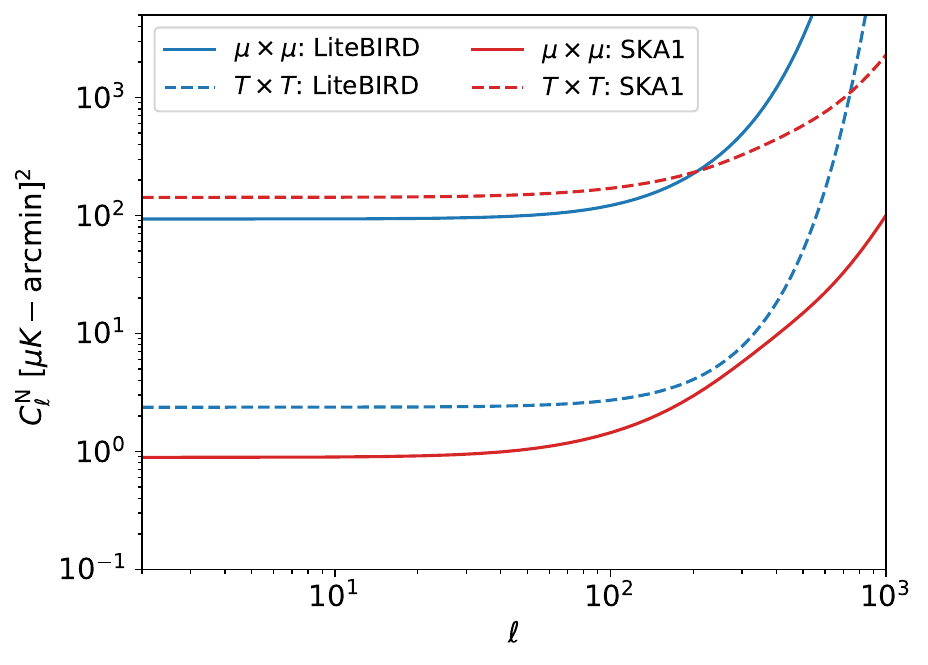}
\caption{$C^{N}_\ell$ for SKA1-MID and LiteBIRD, neglecting foregrounds. Solid lines correspond to $C^{\mu\mu,N}_\ell$, while dashed lines are $C^{TT,N}_\ell$. Despite SKA1-MID's temperature noise being almost two orders of magnitude worse than LiteBIRD's, SKA1-MID has substantially lower noise for $\mu$ owing to its low-frequency coverage. 
\label{fig:mumu}}
\end{figure}

We propose to pair SKA1-MID with a wide survey at traditional CMB frequencies ($\nu \gtrsim20$\;GHz), from which we can obtain signal-dominated maps of $T$ and $E$ anisotropies, and to further enhance foreground subtraction in the $\mu$ map. Given that we are targeting low-$\ell$ ($\ell \lesssim 100$) $\mu$-anisotropies,
sensitivity at large angular scales is more important than angular resolution. For foreground removal, more individual frequency channels are preferred.
One survey that satisfies these criteria is the all-sky survey planned for the space-based telescope LiteBIRD \cite{LiteBIRD:2022cnt}, the primary science goal of which is constraining low-$\ell$ CMB B-mode polarization.
The angular resolution $\theta_i$ and noise level $N^i$ of each frequency band for LiteBIRD is given in Table 1 of \cite{Remazeilles:2021adt}. SKA1-MID's wide survey corresponds to a $f_\text{sky} = 0.48$, which we will limit our forecast of LiteBIRD to.
Fig.~\ref{fig:mumu} shows SKA1-MID and LiteBIRD fill highly complementary roles: LiteBIRD is more sensitive to signals with a blackbody SED, while the low-frequency coverage of SKA1-MID results in much better sensitivity to $\mu$.

\subsection{Calibration}

Precise inter-frequency calibration is needed to extract the very small $\mu$-distortion signal from the much larger $T$ and foreground signals. A small gain mismatch between bands will cause $T\rightarrow\mu$ leakage in the component separation, resulting in a $T \times T$ component to the $\mu \times T$ spectrum. Even if this can be modeled, it will cause excess variance in the $\mu \times T$ spectrum and degrade the constraints on \fnl. This implies that the leaked $T$ should be kept below the level of the noise in the $\mu$ map, which in turn implies that the precision of the relative calibration between bands should be comparable to the map noise divided by the $T$ signal or better. This is satisfied by maps in which the dominant signal is the CMB temperature anisotropy,
which is true for the workhorse bands of LiteBIRD, but is not for SKA. For SKA bands, which are expected to be dominated by synchrotron emission at the multipole ranges of interest, a potential strategy is to first model the synchrotron 
and then calibrate on the CMB using cross-spectra with the CMB-S4 maps. The synchrotron model could be varied, with the calibration effectively marginalized over the uncertainty in the synchrotron model parameters. 
This overall calibration strategy will be tested in upcoming work; for the purposes of this forecast we assume perfect calibration.

\section{Results and Discussion}
\label{sec:results}

\begin{table*}[ht]
\def\arraystretch{0.8}
\setlength{\tabcolsep}{7pt}
\centering
\begin{tabular}{|l|ccc|}
\hline $\sigma(\fnl)$ & \textrm{No foregrounds} & \textrm{Foregrounds} & +\textrm{Moment expansion} \\
\hline LiteBIRD & 91  & 826  & 923   \\
\hline \text{SKA1-MID} & 8 (19/33) &  31 (56/130) & 234 (505/805)   \\
\hline LiteBIRD \& \text{ SKA1-MID} & 6 (13/19) &  23 (61/124) & 92 (141/181)  \\
\text{  + 15-24\;GHz} & 6 (13/18) & 22 (56/100)  & 68 (100/126)  \\
\hline
\end{tabular}
\caption{
Constraints on $\fnl$ from different survey configurations of SKA1-MID and LiteBIRD. We list constraints considering no foregrounds, isotropic foregrounds, and foregrounds with relevant moment expansion terms. Constraints not in parentheses assume $\ell_\textrm{min}=2$, while the ones in parentheses are given for ($\ell_\textrm{min}=50$/$\ell_\textrm{min}=100$)
}
\label{tab:results}
\end{table*}

Table \ref{tab:results} summarizes the forecasted constraints on \fnl\ from SKA1-MID, LiteBIRD, and their combination. In all cases, adding SKA1-MID data improves the \fnl\ constraint over LiteBIRD alone by a factor of at least 10. In the most ideal cases, SKA1-MID alone is nearly as powerful as the combination, but in the most realistic foreground case, the combination is a factor of 3-4 more powerful than SKA1-MID, highlighting the synergy between SKA and traditional CMB experiments.

\begin{figure}[t!]
\centering
\includegraphics[width=1\columnwidth]{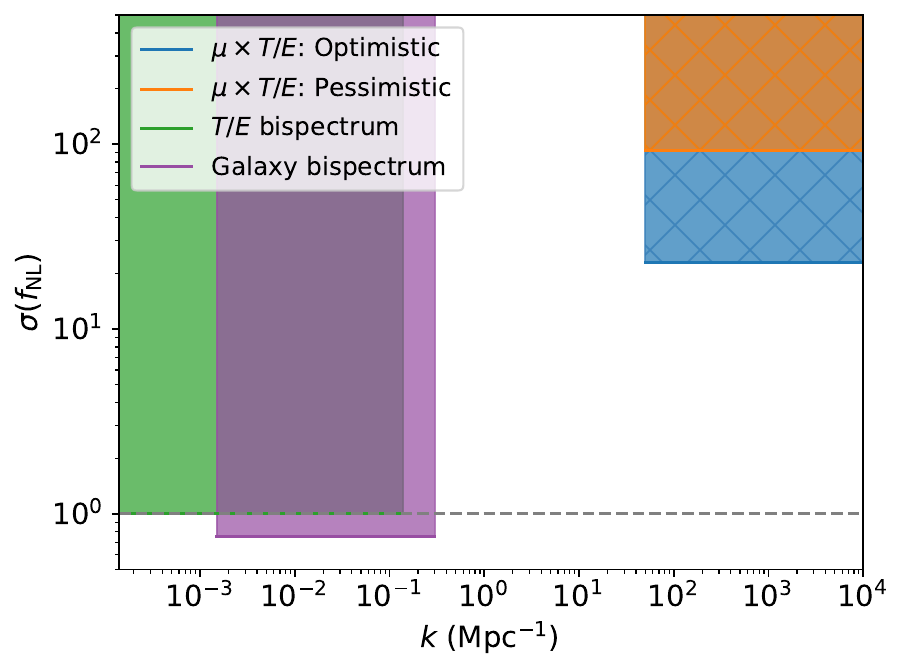}
\caption{Upper bounds on squeezed-limit $\fnl$ at different scales from various observables. These scales are approximations and purely for illustration. CMB bispectrum constraints are from CMB-S4 with scales derived from $\ell_\textrm{min}=2$ and $\ell_\textrm{max}=1000$. Galaxy bispectrum constraints are from SPHEREx \cite{Heinrich:2023qaa} with $k_\textrm{min} \approx 0.0015 \ \textrm{Mpc}^{-1}$  and $k_\textrm{max}=0.3 \ \textrm{Mpc}^{-1}$ \cite{SPHEREx:2014bgr}. For $\mu$-anisotropies from SKA1+LiteBIRD, ``optimistic'' is the second column of Table \ref{tab:results}, while ``pessimistic'' is the third column. For reference, the gray dashed line is $\fnl=1$. Figure inspired by \cite{LoVerde:2007ri,Sabti:2020ser}. 
\label{fig:fnl}}
\end{figure}

In terms of the absolute level of the \fnl\ constraints, constraints approach $\sigma(\fnl)=6$ in the most ideal case, 
comparable to what a cosmic variance-limited measurement of the CMB bispectrum can achieve for large-scale non-Gaussianity, but at much smaller scales ($k\approx740\ \textrm{Mpc}^{-1}$, see Fig.~\ref{fig:fnl}). 

In the most realistic case, combined constraints degrade to $\sigma(\fnl)=92$, which would still be the strongest constraint on non-Gaussianity at such small scales, improving current constraints from $\mu \times T$ \citep{Rotti:2022lvy} by a factor of $\simeq 30$. 

The largest degradation in combined constraints comes from introducing spatially varying foregrounds (specifcally synchrotron---removing dust entirely from the covariance has almost no effect on $\sigma(\fnl)$).
We note that observations from SKA1-LOW can extend coverage down to 50\;MHz, covering the peak of the $\mu$-distortion SED and adding high-signal-to-noise observations of synchrotron. Future work should therefore investigate the potential of using low-$\ell$ measurements from SKA1-LOW.

We have made certain assumptions in this analysis that may be optimistic when compared to real data. We have ignored the effects of instrumental ``$1/f$'' or ``red'' noise (see, e.g., \cite{Harper:2017gln}).
In single-dish mode, instrumental $1/f$ noise will corrupt angular modes with scales larger than $\sim v_\mathrm{scan} / f_\mathrm{knee}$ in the scan direction, where $v_\mathrm{scan}$ is the telescope scanning velocity, and $f_\mathrm{knee}$ is the frequency at which the $1/f$ noise equals the white noise. The SKA precursor experiment MeerKAT has achieved a raw $f_\mathrm{knee} \simeq 0.1 \, \mathrm{Hz}$ \cite{Li:2020bcr}, which would contaminate modes with $\ell \lesssim 50$, for a scan velocity of 1 degree/s.
To account for $1/f$ noise with a plausible range of $f_\mathrm{knee}$ and scan speed values, we include in parentheses in Table \ref{tab:results} results for $\sigma(\fnl)$ with $\ell_\mathrm{min}=50$ and 100 to indicate the degradation.
In addition, we have ignored free-free emission and the ARCADE excess as foregrounds, choices motivated in \S\ref{sec:fg}.
Finally, we also ignored calibration uncertainties, a more robust estimate of which will inform future forecasting for SKA.

There are also aspects of our analysis that might be overly pessimistic. The assumed bands are much wider than SKA1-MID's planned capability of dividing bands into $\Delta\nu \sim 10$ kHz channels. The use of very narrow bands can improve our ability to isolate the $\mu$-anisotropy signal. 
In addition, we are assuming a constant synchrotron amplitude across the Southern sky, using the mean value over the $\fsky = 0.48$ region treated in \cite{Zegeye:2023pbl}. A more careful choice of observing region, and dividing the region into multiple patches, could reduce the impact of synchrotron significantly.

This single worked example of constraining primordial non-Gaussianity through correlations between $\mu$-distortion anisotropy and CMB temperature and polarization anisotropies demonstrates the impressive potential of SKA as a CMB experiment, particularly when combined with a traditional, higher-frequency CMB experiment such as LiteBIRD.
Our forecasts can also be used to constrain other sources of $\mu$-anisotropies, such as modulated thermalization in the $\mu$-era by a long-wavelength curvature mode \cite{Cabass:2018jgj,Zegeye:2021yml} or due to energy injections \cite{Chluba:2022xsd,Chluba:2022efq,Kite:2022eye}. As seen in Fig. \ref{fig:fnl}, our constraints on $\fnl$ place bounds on primordial non-Gaussianity at scales inaccessible to other cosmological observables. If squeezed-limit non-Gaussianity grows at smaller scales (i.e has a blue-tilted spectrum), it results in a proportional improvement on our constraints on $\fnl$.

Future forecasts should expand on this single example to other CMB science cases, including $B$-modes from primordial gravitational waves, and also include the even more impressive raw sensitivity of the planned SKA2 upgrade.

\acknowledgments

We would like to thank Peter Adshead, Darcy Barron, Ritoban Basu Thakur, Federico Bianchini, Daniel Grin, Gilbert Holder, Wayne Hu, Daan Meerburg, Giorgio Orlando, Tristan Smith, Subodh Patil, and Andrea Ravenni for all the useful discussions throughout this project's journey. 

D.Z.~acknowledges support from National Science Foundation award AST-2240374 and the National Science Foundation Graduate Research Fellowship Program under Grant No. DGE1746045 
T.C.~acknowledges support from National Science Foundation award OPP-1852617.
J.C. was supported by the ERC Consolidator Grant {\it CMBSPEC} (No.~725456) and the Royal Society as a Royal Society University Research Fellow at the University of Manchester, UK (No.~URF/R/191023).
M.R.~acknowledges financial support provided by the project refs. PID2022-139223OB-C21 and PID2022-140670NA-I00 funded by the Spanish MCIN/AEI/10.13039/501100011033/FEDER, UE.

\bibliography{main}

\end{document}